# Reaction-network reasoning with frontier models for experimentally confirmed catalyst-selectivity hypotheses

Sutanay Choudhury[1*], Anwesha Banerjee[2], Udishnu Sanyal[1], Jorin Dawidowicz[3], Chiezugolum Ijeoma Odilinye[3], Jesun Firoz[1], Liney Arnadottir[1,3], Simone Raugei[1], Johannes Lercher[4], Arnab Dutta[2*]

[1] *Pacific Northwest National Laboratory, USA*
[2] *Indian Institute of Technology, Bombay, India*
[3] *Oregon State University, USA*
[4] *Technical University of Munich, Germany*

* Corresponding authors:
Sutanay Choudhury (Sutanay.Choudhury@pnnl.gov);
Arnab Dutta (arnabdutta@chem.iitb.ac.in).

## Abstract

Catalysts are essential for sustainable chemical manufacturing, yet discovering novel architectures remains a bottleneck dominated by trial-and-error experimentation and computationally intensive screening. In complex reactions like electrochemical carbon dioxide reduction, product selectivity is governed by dynamic interfacial, electrolyte, and potential factors and kinetic pathway competition. Conventional descriptor-based machine learning and computational potentials struggle to resolve these mechanistic branch points, primarily relying on static ground-state descriptors or bulk structural correlations rather than end-to-end topological pathway analysis. Here, we show that frontier language models, when strictly constrained to reason over explicit reaction networks, can discover novel catalysts by identifying the physical levers that govern pathway competition. We developed a human–AI co-thinking framework that enforces network invariance to extract testable hypotheses from complex chemical graphs. Applied to CO2 electroreduction, the framework identified ketene desorption and hydroxide capture as the acetate-forming pathway, and predicted a distinct CO*-CH2* coupling route to ketene. By isolating actionable control levers, specifically local alkalinity, controlled iron incorporation, and restricted interfacial proton-donor accessibility, the framework guided the prospective synthesis of a copper-iron oxide catalyst demonstrating a threefold increase in acetate selectivity over matched Cu-rich baselines. This mechanism-guided reasoning architecture shifts the computational paradigm from retrospective statistical prediction to forward-looking hypothesis generation, providing a broadly applicable blueprint for mechanism-guided materials discovery.

## Introduction

From ammonia synthesis[1] to polymer production[2], catalysts have continuously transformed human civilization. However, the critical challenge on sustainable manufacturing[3] still hinges on effective control of carbon–carbon coupling i.e., steering the microscopic competition among divergent reaction pathways toward high selectivity. In such systems, several pathways share the same intermediates and diverge at only one or two elementary branch points[11-13]. Consequently, whether a catalyst surface produces to a valuable oxygenate (acetate) or an easier hydrocarbon (ethylene) is set by local environmental factors, such as interfacial acidity and the surrounding electrolyte[4-6].

For such systems, a useful hypothesis must be more than a plausible explanation but provide falsifiable mechanistic reasoning. It must identify a specific mechanistic branch point, the physical quantity that governs it, and an experimental observation capable of verifying the prediction. The computational tools that currently dominate catalyst discovery fail to resolve these critical branch points. First-principles theory is time consuming, in particle for branching reaction networks on complex transition-metal oxides like copper-iron[7]. Machine-learning potentials built to accelerate these calculations and descriptor-based machine learning predicts activity well where data are dense, but lacks the representation to capture the environmental descriptors that actually govern selectivity[8-10].

We overcome these computational blind spots by forcing a frontier large language model (LLM) to reason exclusively over an explicit reaction network[14], a design principle we term *network invariance*. By instantiating this directed graph of species and elementary steps as a digital twin, the agent is forced to condition every proposed hypothesis on a shared topological map, ensuring its reasoning is a function of the physical chemistry rather than the model's idiosyncratic priors. Unlike standard scientific agents[15-16] that return broad propositions prone to drift across conversational threads[17-19], this constraint ensures that the "co-thinking" between human and AI remains rigorously coherent across iterations. Rather than attempting to correlate bulk catalyst features directly to a final product yield, the model is anchored to defined chemical transitions, forcing it to identify the specific mechanistic branch points and localized descriptors such as interfacial acidity that dictate selectivity.

To execute this principle, we developed CoThinker, establishing a methodological foundation for mechanistic reasoning as researchers push into novel chemical spaces. The resulting discovery of a 1:1 Cu–Fe oxide catalyst serves not merely as a performance milestone, but as physical proof of this design principle. The catalyst increases acetate selectivity approximately threefold relative to matched Cu-rich baselines. Although molecularly engineered systems such as dendrimer-functionalized Cu can achieve high acetate Faradaic efficiencies and partial current densities[23], they rely on complex surface functionalization to construct the active microenvironment. In contrast, the CoThinker-derived catalyst achieves this selectivity through a minimal, earth-abundant mixed-oxide architecture, offering a potentially more scalable route for mechanism-guided catalyst discovery. Crucially, because the agent outputs physical logic rather than purely correlative predictions, we uniquely validate its hypotheses across both domains: confirming the mechanisms theoretically via first-principles computation and proving the outcomes physically through forward experimentation. Unlike established approaches that optimize design spaces via active learning without explicit physical models[20], or that validate language-model proposals strictly against retrospective computational potentials[21], this framework shifts the computational objective from predicting a singular winning material to providing a generalizable blueprint for deriving falsifiable, per-step reaction mechanisms.

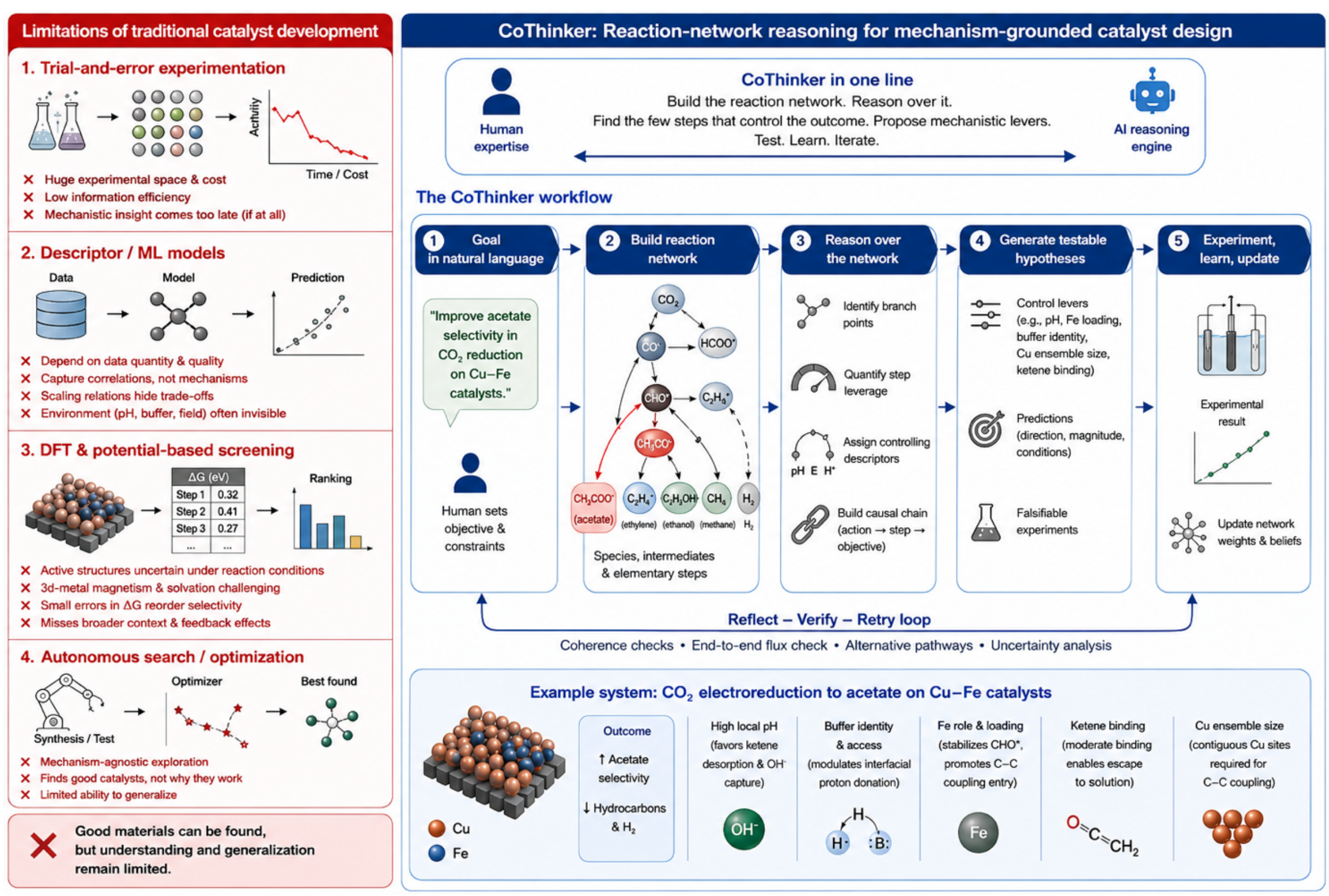

**Figure 1 | Reaction-network reasoning for mechanism-guided catalyst discovery.** Left, conventional strategies including descriptor-based machine learning, computational screening, and autonomous optimization efficiently explore design spaces but lack explicit mechanistic reasoning. Right, the CoThinker workflow translates a catalytic objective into an explicit reaction network. By evaluating competitive branch points and enforcing end-to-end pathway coherence through a structured reflect-verify-retry loop, the framework extracts experimentally testable mechanistic hypotheses. Applied to the electrochemical reduction of $CO_2$ to acetate, this hierarchical planning yields actionable control levers such as local pH, proton accessibility, and Fe incorporation to guide prospective experimental validation.

## Results

The CoThinker framework executes a hierarchical planning workflow[34] grounded in the physical state space of the reaction (Fig. 1). The process initiates when a human expert defines the catalytic objective, which the system formally translates into a mechanistic optimization problem. To establish the topological boundaries of the reasoning task, the platform first constructs the complete reaction network of species and elementary steps. Rather than relying on unconstrained text generation, the reasoning engine navigates this network to identify competitive branch points[14], quantify step leverage, and assign controlling physical descriptors. Before generating a final hypothesis, the system employs a structured self-critique loop[32,33] for evaluating decisions that may be locally optimal but globally inconsistent[30,31]. The output of this constrained planning phase is a prioritized set of falsifiable control levers, which are subsequently passed to the users for prospective experimental execution.

Crucially, every mechanistic claim reported herein is derived solely by a frontier reasoning model (OpenAI GPT-5.4[39]) operating over its encoded literature knowledge. Present CoThinker architecture deliberately bypasses explicit computational chemistry calculations using ab-initio simulations or neural network potentials (NNPs). While foundational NNPs[24,25] offer rapid energy screening, they suffer from severe out-of-distribution errors at complex electrochemical interfaces[26,27] and are predominantly trained on ground states rather than the transition-state kinetics required to resolve pathway competition[28,29]. Although language models possess inherent inferential uncertainties, intentionally restricting the framework purely to literature-grounded logical deduction in this study prevents the compounding of errors that arises from coupling language generation with unstable numerical extrapolations.

### *Reaction-network reasoning identifies mechanistic determinants of selectivity*

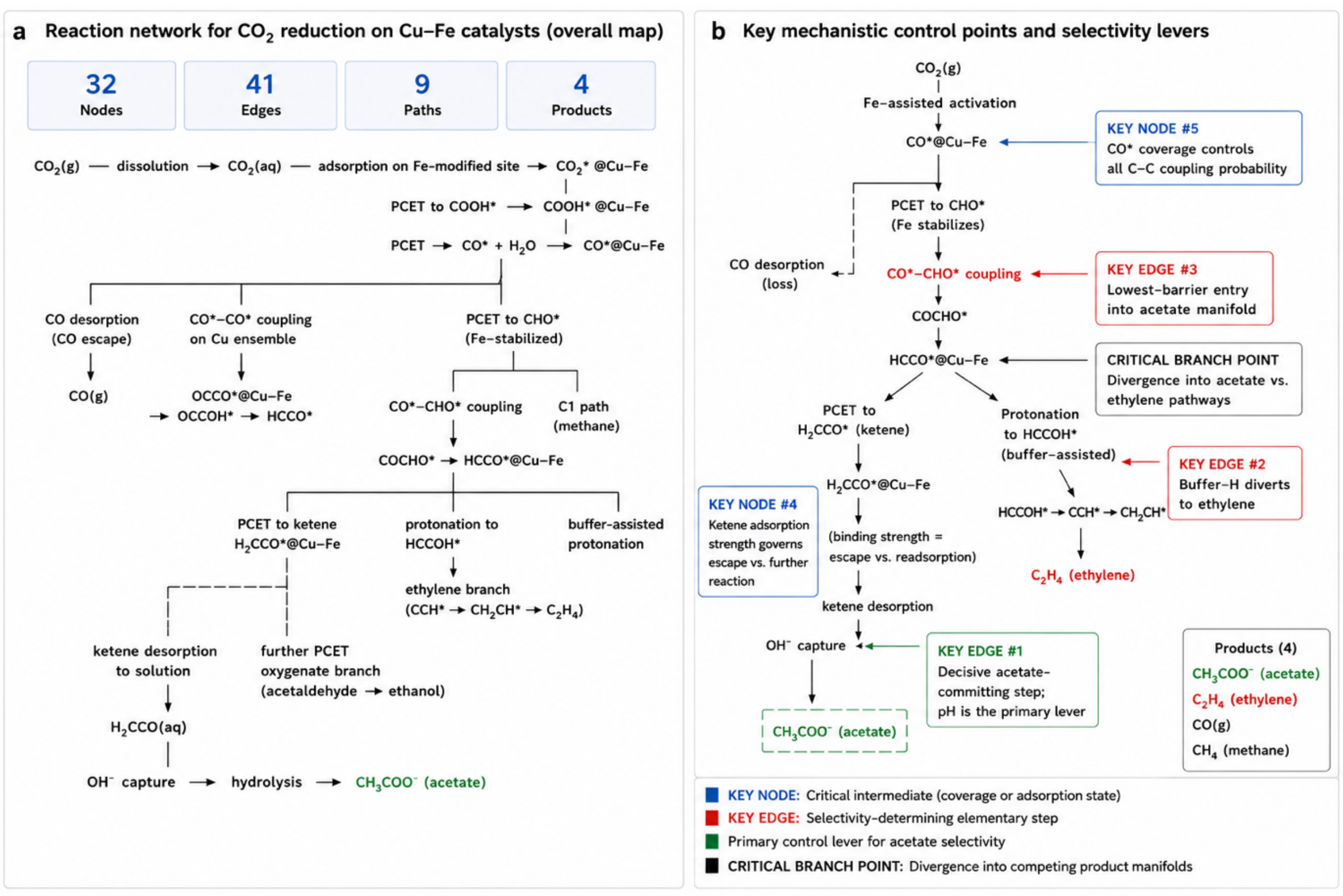

**Figure 2 | Reaction-network analysis identifies the mechanistic determinants of acetate formation on Cu–Fe catalysts. a**, Complete reaction network linking $CO_2$ activation to competing $C_1$ and $C_2$ products through interconnected elementary reaction pathways. The network consists of 32 intermediates, 41 elementary reaction steps, nine reaction pathways, and four product manifolds, enabling explicit representation of pathway competition. **b**, Mechanistic hotspots extracted from the reaction network. Five high-leverage control elements, including CO* coverage, CO*–CHO* coupling, ketene adsorption strength, buffer-assisted protonation, and hydroxide-assisted ketene capture, govern the distribution of reaction flux, with hydroxide-assisted ketene capture identified as the acetate-forming pathway[13].

Establishing the mechanistic scope of this reasoning task requires constructing an explicit reaction network linking $CO_2$ activation to the principal $C_1$ and $C_2$ products (Fig. 2). The initial network of 32 chemical species and 38 elementary transformations was iteratively expanded by the reasoning engine to incorporate missing competitive pathways, such as the buffer-assisted protonation of the $HCCO^*$ intermediate. The resulting complete reaction graph comprises 41 elementary steps spanning the acetate, ethylene, ethanol, methane, and formate manifolds, enabling the evaluation of selectivity as a competition among complete end-to-end pathways rather than isolated reactions. Evaluating this complete network reveals that only a small subset of elementary transformations, specifically five mechanistic hotspots exerts dominant control over product selectivity (Fig. 2 and Table 1). Table 1 shows the exact reasoning summary produced from the CoThinker run.

**Table 1 | Ranked mechanistic hot-spots for Cu–Fe $CO_2$-to-acetate.**

| # | Type | Element | Descriptors (critical in bold) | Mechanistic rationale |
|---|---|---|---|---|
| 1 | edge | $OH^-$-assisted ketene desorption & capture ($H_2CCO^*$ → $H_2CCO\cdots OH^-$) | **activation barrier**; **local $OH^-$ activity**; **transition-state energy**; coverage; solvation | Decisive acetate-forming step: once ketene is pulled off the surface and trapped by $OH^-$, flux locks into acetate. Strongest handle for pH. |
| 2 | edge | Buffer-assisted protonation ($HCCO^*\cdots$Buffer-H → $HCCOH^*$) | **proton-donor p$K_a$**; **interfacial buffer concentration**; steric hindrance | Key electrolyte/buffer control: accelerates the competing ethylene branch from the shared $HCCO^*$ intermediate. |
| 3 | edge | $CHO^*$–$CO^*$ coupling ($CHO^*$ + $CO^*$ → $COCHO^*$) | coverage; d-band centre; coordination number | Lowest-barrier C–C entry into the acetate manifold and the clearest role for Fe: Fe stabilisation of $CHO^*$ makes coupling competitive against CO loss and C1 hydrogenation. |
| 4 | node | Adsorbed ketene, $H_2CCO^*$@Cu–Fe | **binding energy**; d-band centre; **coordination number**; **interfacial field** | Central branch-point intermediate: binding strength sets desorption to solution (acetate) versus further surface reduction (ethylene/ethanol). |
| 5 | node | $CO^*$ coverage on Cu–Fe | **local $OH^-$ activity**; **transition-state energy**; coverage; solvation | Controls downstream C–C coupling probability; Fe acts indirectly by tuning how much activated CO is supplied near Cu coupling sites. |

The highest-ranked hotspot was the hydroxide-assisted desorption and solution-phase capture of ketene ($H_2CCO^*$), which the framework identified as the acetate-forming pathway[13].

Because this reaction irreversibly commits the reaction flux toward acetate formation, the activation barrier for ketene desorption, together with the local hydroxide activity, emerged as the strongest mechanistic handle controlling selectivity. The second hotspot corresponded to the protonation of the shared $HCCO^*$ intermediate by electrolyte-derived proton donors[38]. This elementary step diverts reaction flux from the acetate manifold into the competing ethylene pathway, thereby identifying buffer identity and proton accessibility as critical determinants of selectivity. Third, the framework identified $CHO^* - CO^*$ coupling as the lowest-barrier entry into the acetate-forming pathway and the most probable point at which Fe incorporation influences catalytic behaviour. Rather than acting as an independent active site, Fe was predicted to function as an electronic modifier that stabilizes CHO-containing intermediates sufficiently to promote carbon-carbon coupling while preserving the contiguous Cu ensembles required[5,37] for subsequent reduction chemistry. Fourth, adsorbed ketene emerged as the principal branch-point intermediate controlling the competition between solution-phase acetate formation and continued surface reduction toward ethylene or ethanol. Its adsorption strength, therefore, represents a key kinetic descriptor governing product selectivity. Finally, the surface coverage of $CO^*$ was identified as the principal descriptor[13] controlling the probability of downstream carbon-carbon coupling. Rather than directly participating in C–C bond formation, Fe was predicted to influence this variable indirectly by modifying the local availability of activated CO species adjacent to Cu-rich ensembles.

***Iterative reasoning eliminates locally optimal but globally inconsistent designs***

Because local energetic improvements to individual elementary steps frequently fail to shift macroscopic selectivity[30,31], true catalyst optimization requires evaluating end-to-end pathway coherence. CoThinker enforces this requirement through a structured reflect–verify–retry loop that subjects every generated hypothesis to explicit falsification against the complete reaction network (Fig. 3). Rather than accepting the first chemically plausible output, the framework systematically evaluates whether a proposed modification improves the overall reaction flux, maintains internal consistency across competing branches, and remains physically viable under realistic electrochemical conditions. Candidate architectures that fail any of these criteria are discarded and replaced.

As captured in the model's raw reasoning logs (Fig. 3), initial reasoning cycles frequently generated candidate designs that seek to optimize local kinetics but failed global network constraints. For example, early proposals incorporating subsurface Fe or conformal fluoropolymer overlayers would effectively lower the activation barrier for initial carbon–carbon coupling or improved interfacial transport. However, network-wide verification rejected these designs because they either violated scaling-relation constraints[35], simultaneously lowering barriers for competing ethylene and ethanol pathways, or failed to resolve the unchanged $CO^* - CHO^*$ coupling bottleneck upstream[36]. After executing these structured falsification cycles, the reasoning engine finally converge to an internally consistent architecture: Cu–Fe oxide interfaces decorated with sparsely distributed hydrophobic ion-conducting domains. Positioned away from contiguous Cu ensembles, this accepted configuration cooperatively addresses the global network constraints by simultaneously stabilizing intermediate coupling, supporting local alkalinity for irreversible ketene capture, and selectively suppressing protonation toward competing pathways without introducing mass-transfer resistance.

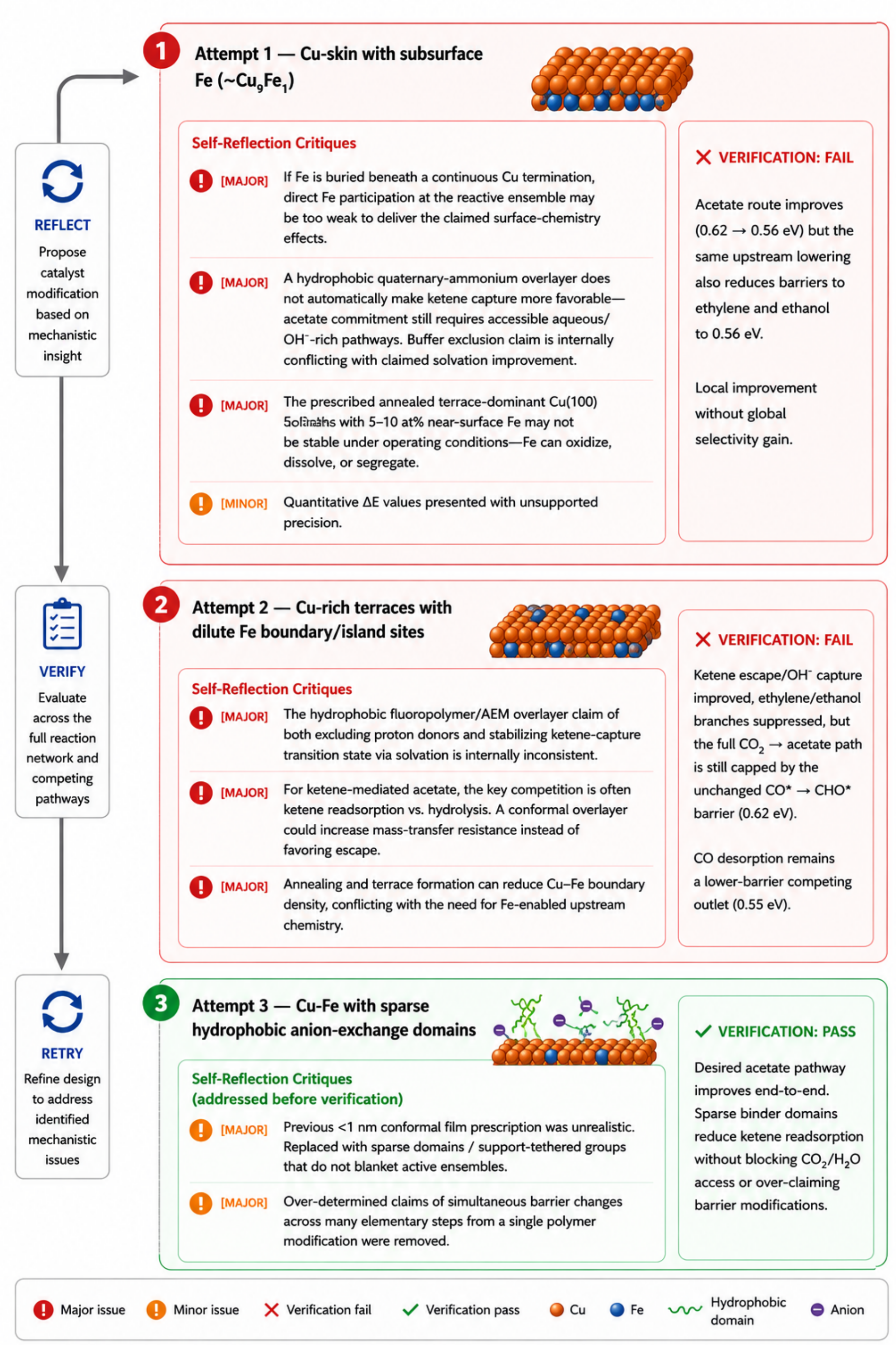

**Figure 3 | Iterative mechanistic refinement of catalyst-design hypotheses.** Candidate catalyst designs generated by CoThinker are evaluated using a reflect-verify-retry workflow. The first two designs are rejected because local improvements in individual reaction steps do not translate into enhanced overall acetate selectivity across the complete reaction network. The final design resolves these mechanistic inconsistencies by combining Cu-Fe interfaces with sparse hydrophobic anion-exchange domains, promoting ketene desorption while suppressing competing protonation pathways and preserving contiguous Cu ensembles for C–C coupling. The accepted catalyst is selected on the basis of end-to-end mechanistic consistency rather than isolated energetic improvements.

***Translating network constraints into physical control levers***

The outcome of the iterative refinement loop is not a single optimized catalyst composition, but a mechanistically constrained set of experimentally testable design principles (Table 2). Having survived the network-wide falsification process, the accepted architecture, "Cu–Fe interfaces decorated with sparse hydrophobic anion-exchange domains", was further reasoned in depth to isolate its functional components. Because navigating the unknown catalyst design space carries high computational uncertainty, the framework decomposes this architecture into independent, literature-grounded control variables. By stepping into this unknown space one high-confidence, experimentally falsifiable action at a time, the framework translates a complex structural hypothesis into a directed laboratory campaign.

For the Cu-Fe $CO_2$ reduction system, the accepted catalyst design was distilled into five mechanistic control variables governing acetate selectivity (**Table 2**). Three of these variables, local pH, proton-donor accessibility, and Fe incorporation, are directly tunable experimentally, whereas ketene binding strength and Cu ensemble size emerge as intrinsic constraints that define the feasible catalyst design space. Together, these variables provide a mechanistic map linking catalyst composition, reaction environment, and product selectivity.

**Table 2 | Selectivity control levers distilled for Cu-Fe $CO_2$-to-acetate.**

| Lever | Mechanism | Design implication |
|---|---|---|
| Local pH ($OH^-$) | Promotes concerted ketene desorption and hydroxide capture- the acetate forming pathway | Operate at high local pH; choose a morphology generating local alkalinity at the surface |
| Buffer identity & access | Interfacial proton donors accelerate the protonation that diverts flux toward ethylene | Minimise interfacial proton-donor availability; a sparse hydrophobic binder restricts proton-donor approach |
| Fe role & loading | Fe stabilises CHO upstream, enabling CO–CHO coupling into the acetate manifold | Fe as an electronic modifier near Cu ensembles; retain contiguous Cu domains |
| Ketene binding | Adsorbed-ketene binding strength controls desorption-to-solution versus surface reduction | Cu-like (moderate) binding preferred; avoid over-binding from excess Fe at active sites |
| Cu ensemble size | Contiguous Cu sites are required for C–C coupling | Maintain Cu-rich domains; do not atomically dilute Cu into an Fe matrix |

The highest-priority design principles concern the local reaction environment and electrolyte composition. CoThinker identified hydroxide-assisted desorption and solution-phase capture of ketene as the acetate forming pathway[13], dictating that any architecture capable of generating local alkalinity should favour acetate formation. This naturally establishes local pH as the primary experimentally controllable design variable. Conversely, because protonation[38] of the shared $HCCO^*$ intermediate serves as the principal entry point into the competing ethylene pathway, proton-donor accessibility at the catalyst–electrolyte interface must be minimized. Given these constraints and the goal of hypothesis generation, the reasoning model proposed sparse hydrophobic ion-conducting domains. This specific architecture explicitly avoids conformal overlayers that overestimate barrier modifications, instead selectively regulating interfacial proton access while maintaining adequate transport of $CO_2$, water, and escaping ketene.
The third experimentally accessible control variable concerns the role of Fe incorporation. The framework consistently predicts that Fe functions as an electronic modifier operating upstream of carbon-carbon coupling, where stabilization of CHO-containing intermediates increases the probability of productive $CO^* - CHO^*$ coupling. However, because contiguous Cu ensembles emerge as a structural prerequisite for efficient carbon–carbon coupling[37], excessive Fe incorporation is explicitly predicted to fragment these domains and reduce overall coupling probability. The resulting design principle therefore emphasizes controlled Fe incorporation adjacent to Cu-rich domains rather than the maximization of Fe content.

The remaining mechanistic variables define intrinsic structural boundary conditions rather than independently tunable parameters. Most notably, adsorbed ketene acts as the central branch-point intermediate, meaning its adsorption strength must remain sufficiently weak to permit desorption into solution while remaining strong enough to enable productive surface chemistry. Together with the strict structural requirement for contiguous Cu domains, these variables establish the mechanistic limits within which the experimentally adjustable parameters - local pH, electrolyte composition, and Fe loading can be systematically varied.

***Forward experimental and computational validation of predicted mechanistic control levers***

To prospectively test the AI-derived mechanistic hypotheses, a series of Cu–Fe oxide catalysts with varying stoichiometries (Cu:Fe ratios of 4:1, 3:1, 2:1, and 1:1) was synthesized and evaluated independently of the reasoning process. By fixing the mechanistic predictions prior to synthesis, the subsequent electrochemical evaluation serves as a strict forward validation of the predicted control variables rather than a post hoc rationalization of empirical performance.

Consistent with the identification of hydroxide-assisted ketene capture as the acetate-forming pathway[13], elevating the local electrolyte pH systematically shifted reaction flux toward acetate. Increasing the initial bicarbonate electrolyte pH from 6.8 to 7.5 enhanced both acetate production rates and Faradaic efficiencies across all investigated catalyst compositions. The 1:1 Cu–Fe catalyst exhibited the most pronounced response, directly supporting the theoretical hypothesis that local hydroxide activity fundamentally governs the final branch point independently of bulk catalyst composition.

Systematic variation of the Cu:Fe ratio confirmed the framework's prediction that iron functions as a tunable electronic modifier rather than an independent acetate-producing active site. While the Cu-rich 4:1 catalyst predominantly produced formate, progressive iron incorporation smoothly redirected reaction flux into the acetate manifold, culminating in peak acetate selectivity for the 1:1 composition. Crucially, high-resolution transmission electron microscopy (TEM) and elemental mapping revealed that the Cu component maintains contiguous domains across the compositional series, while iron oxide segregates to peripheral regions. This physical architecture perfectly mirrors the network-derived structural requirement: iron modifies the electronic landscape to promote upstream coupling while strictly preserving the contiguous copper ensembles required for subsequent carbon–carbon bond formation.

Modulating the interfacial proton-donor accessibility via electrolyte substitution validated the predicted kinetic competition at the shared $HCCO^*$ intermediate. Electrolysis performed across bicarbonate, phosphate, and chloride environments revealed stark divergences in acetate selectivity. While bicarbonate electrolytes supported sustained acetate production, phosphate-buffered systems—which provide higher interfacial proton availability—exhibited suppressed acetate yields and progressive catalyst passivation. This confirms the system's prediction that local proton-donor accessibility explicitly diverts flux away from the target product toward competing pathways.

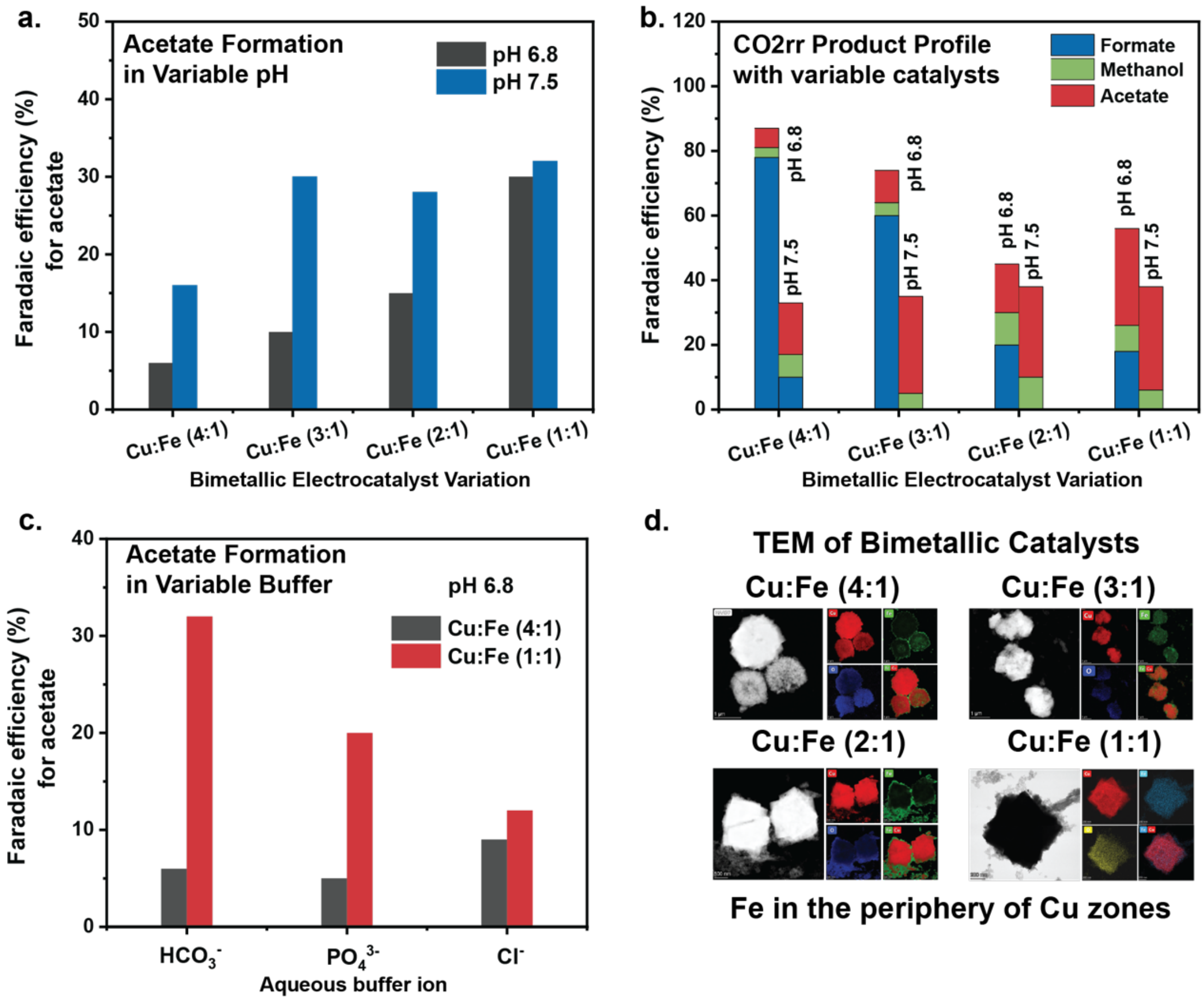

**Figure 5 | Prospective experimental validation of the mechanistic control variables governing acetate formation on Cu–Fe oxide catalysts.** **a**, Effect of electrolyte pH on acetate selectivity across the Cu–Fe catalyst series. Increasing the electrolyte pH from 6.8 to 7.5 systematically enhances the Faradaic efficiency for acetate formation, consistent with the predicted role of hydroxide-assisted ketene capture as the acetate forming pathway. **b**, Product distribution as a function of Cu composition, showing the evolution of acetate, methanol, and formate selectivity. Progressive incorporation of Fe redirects reaction flux toward acetate formation, with the CuFe (1:1) catalyst exhibiting the highest acetate selectivity. **c**, Influence of electrolyte identity on acetate production for CuFe (4:1) and CuFe (1:1) catalysts, demonstrating the role of proton-donor accessibility and interfacial reaction environment in governing selectivity. **d**, Representative HAADF-STEM images and elemental

maps of the Cu–Fe catalyst series. The Cu component forms contiguous domains, whereas Fe oxide preferentially occupies peripheral regions, supporting the predicted requirement for contiguous Cu ensembles and positioning Fe as an electronic modifier rather than an isolated active site.

Finally, the intrinsic structural constraint requiring weakened ketene binding to promote desorption-to-solution was validated computationally using Universal Machine-learned Potentials (UMA[24]). Calculating the adsorption energy of isolated Fe substituted into a CuO surface demonstrated a distinct lowering of the ketene adsorption energy from -0.61 eV to -0.48 eV. This supports the hypothesis that Fe destabilizes the bound intermediate to facilitate solution-phase capture. While comparative calculations substituting $Fe_2O_3$-like clusters exhibited strong ketene binding, the highly reducing potentials inherent to the cathodic operating conditions suggest the metal-terminated surface represents the physically realistic active state. Thus, targeted computational analysis confirms the final intrinsic lever.

## Discussion

Catalyst discovery has traditionally progressed through an iterative dialogue between mechanistic intuition and experimental validation. While computational chemistry[24,25] and autonomous laboratories[22] have accelerated the execution of these cycles, they generally address either numerical prediction or objective-function optimization rather than scientific reasoning itself. CoThinker introduces a complementary paradigm: it breaks the traditional barriers of knowledge acquisition by shifting the computational objective from retrospective mechanism clarification to forward-looking, mechanism-centred reasoning. By forcing a frontier language model to evaluate the underlying chemistry, the framework helps scientists navigate complex reaction networks faster and more rigorously, extending the role of artificial intelligence from simply accelerating experimentation to defining the precise scientific questions that experiments must answer.

The central architectural advance of this framework is the introduction of the explicit reaction network as the core abstraction for machine reasoning. Unconstrained language models are prone to contextual drift[17] over chat threads. CoThinker neutralizes this weakness by enforcing network invariance, anchoring every AI-generated proposition to a shared, physically coherent topological map. By translating the domain-expert mental model into a computable directed graph, the system ensures the human–AI co-thinking remains rigorously grounded and logically consistent across iterations.

This topological grounding naturally addresses the high uncertainty inherent in frontier chemical exploration. Conventional descriptor-based machine learning[10] often fails in these spaces because models trained on bulk composition or static ground-state descriptors lack the representation to capture the highly localized, dynamic environmental variables that govern selectivity[8,9]. Rather than blindly optimizing a latent space, CoThinker decomposes the complex state space into a compact set of independent, high-confidence control levers, such as local pH or proton-donor accessibility, that isolate specific mechanistic branch points. By associating each hotspot with a falsifiable physical descriptor, the framework converts a high-dimensional design space into a sequence of high-confidence, independently testable hypotheses, narrowing the space of viable designs one directed experiment at a time.

The experimental validation of the Cu–Fe catalyst serves as physical proof of this design principle, yielding a threefold increase in acetate selectivity over matched Cu-rich baselines while maintaining comparable overall conversion. The system recovers established physical truths while prioritising specific mechanistic levers. The catalyst passivation observed in chloride electrolytes, for example, is consistent with classical halide poisoning, supporting the network's kinetic constraints. The framework also identifies a distinct acetate-committing route in which CO* couples with a surface methylene (CO*–CH2*) to form ketene, in contrast to the CO-dimerization route to ketene described previously. Because these control levers were fixed prior to synthesis, the outcomes validate the predictions prospectively rather than by post hoc rationalisation, indicating that the framework directed reaction flux away from C1 pathways largely by design.

The current framework deliberately isolates language-model reasoning from explicit computational chemistry calculations to strictly prevent error compounding. While neural network potentials (NNPs) offer rapid energy evaluations, they currently suffer from severe out-of-distribution extrapolation errors at complex, multi-component electrochemical interfaces. Furthermore, current foundation models lack the architecture to reliably evaluate the transition-state structures strictly required to resolve kinetic pathway competition. Future iterations must bridge this gap by integrating NNPs via active learning to manage distribution shifts, deploying advanced architectures capable of modelling transition states, and establishing rigorous uncertainty quantification to support a fully closed-loop discovery system.

## Acknowledgement

This work was supported in part by the U.S. Department of Energy (DOE), Office of Science, Office of Basic Energy Sciences, the Division of Chemical Sciences, Geosciences, and Biosciences. Pacific Northwest National Laboratory (PNNL) is a multiprogram national laboratory operated for the DOE by Battelle Memorial Institute under Contract No. DE-AC05-76RL01830. A.D. and A.B. acknowledge the support of Indian Institute of Technology, Bombay for providing the facilities during this research.

## Author Contribution

**S.C.** designed the CoThinker framework. **S.C., A.D., and U.S.** designed the study. **J.F., C.I.O., J.D., and L.A.** performed the neural network potential-based computational modeling. **A.B.** performed all experimental studies. **S.C., A.D., U.S., and L.A.** wrote the manuscript. **J.L.** and **S.R.** provided scientific interpretation and guidance on the analysis. All authors discussed the results and commented on the manuscript.

## Methods

### CoThinker framework

CoThinker is a mechanistically guided reasoning framework designed to support catalyst discovery through structured human–AI collaboration. Rather than directly predicting catalyst compositions from data, the framework constructs an explicit reaction network from a user-defined catalytic objective and reasons over competing elementary reaction pathways to identify mechanistic control variables governing catalytic selectivity.

The workflow consists of four sequential stages: (i) problem formulation, in which the desired catalytic objective and competing products are translated into a structured optimization problem; (ii) reaction-network construction, where intermediates and elementary reaction steps are assembled from literature-derived mechanistic knowledge; (iii) iterative mechanistic reasoning, which identifies selectivity-determining branch points and experimentally controllable descriptors; and (iv) mechanistic verification, where proposed catalyst modifications are evaluated through repeated reflect–verify–retry cycles before being advanced for experimental validation.

### Construction of the reaction network

For electrochemical $CO_2$ reduction, the reaction network was initialized using established mechanistic pathways reported for Cu-based catalysts. Literature-derived elementary reaction steps, adsorbed intermediates, and competing product pathways were assembled into a directed reaction graph. The network was automatically expanded by recursively identifying missing intermediates and competing reaction branches until all experimentally relevant products were connected through complete reaction pathways.

Each node represents a chemically distinct adsorbed or solution-phase intermediate, while each edge corresponds to a single elementary reaction step, including proton-coupled electron transfer (PCET), adsorption, desorption, surface coupling, hydrolysis, and solution-phase reactions. The final reaction graph contained 32 intermediates connected by 41 elementary reaction steps, spanning nine complete reaction pathways leading to four principal product manifolds.

### Mechanistic reasoning and iterative refinement

Mechanistic reasoning was performed over the complete reaction network rather than individual catalyst descriptors. Each elementary reaction step was evaluated according to its influence on reaction flux, competition with alternative pathways, and contribution to overall product selectivity. The framework ranked mechanistic hotspots based on their influence on branching behaviour within the reaction network and identified experimentally accessible control variables.

Candidate catalyst designs generated during reasoning were subjected to an iterative reflect–verify–retry procedure. During each iteration, proposed catalyst modifications were evaluated for (i) mechanistic consistency across the complete reaction network, (ii) pathway completeness, (iii) compatibility with competing reaction branches, and (iv) experimental feasibility. Designs exhibiting local energetic improvements without improving end-to-end pathway selectivity were rejected and replaced through subsequent reasoning iterations. The reasoning process continued until a catalyst design satisfied all verification criteria or the predefined iteration limit was reached. The Cu–Fe $CO_2$ reduction study required three refinement iterations and twenty model calls before convergence to the final mechanistically consistent catalyst hypothesis.

**Catalyst synthesis**

A series of Cu–Fe oxide catalysts with Cu:Fe molar ratios of 4:1, 3:1, 2:1, and 1:1 was synthesized using a modified co-precipitation method. Appropriate amounts of copper and iron precursor salts were dissolved in deionized water and mixed under continuous stirring. Controlled addition of alkaline solution induced simultaneous precipitation of mixed hydroxide precursors, which were aged, washed thoroughly with water and ethanol, and dried under vacuum. The resulting powders were calcined under air to obtain mixed Cu–Fe oxide catalysts. Detailed synthesis procedures and precursor compositions are provided in the Supplementary Information.

**Structural characterization**

Powder X-ray diffraction (XRD) was used to determine crystalline phases. Surface morphology and catalyst microstructure were examined using field-emission scanning electron microscopy (FESEM) and transmission electron microscopy (TEM). High-resolution TEM and elemental mapping were employed to determine the spatial distribution of Cu and Fe within the mixed oxide catalysts. The oxidation states and surface chemical environments of Cu and Fe were analysed using X-ray photoelectron spectroscopy (XPS).

The TEM analyses showed that Cu-rich regions remained as contiguous ensembles throughout the Cu–Fe catalyst series, whereas Fe oxide preferentially formed peripheral domains surrounding the Cu-rich regions. This structural arrangement was compared directly with the mechanistic design principles identified by the reaction-network analysis.

**Electrochemical $CO_2$ reduction**

Electrochemical $CO_2$ reduction experiments were performed in a gas-tight three-electrode electrochemical cell using a $CO_2$-saturated aqueous electrolyte. Catalyst inks were prepared by dispersing catalyst powder in a mixture of solvent and binder and deposited onto carbon paper to form the working electrode. A platinum mesh and Ag/AgCl electrode served as the counter and reference electrodes, respectively. All reported potentials were converted to the reversible hydrogen electrode (RHE) scale.

Electrochemical measurements were conducted using $CO_2$-saturated electrolytes of varying pH and electrolyte composition to evaluate the mechanistic predictions generated by CoThinker. Current densities were corrected for solution resistance where appropriate.

**Product analysis**

Gaseous products were analysed by online gas chromatography equipped with thermal conductivity and flame ionization detectors. Liquid products were quantified using proton nuclear magnetic resonance spectroscopy ($^1$H NMR) with an internal standard following electrolysis.

Faradaic efficiencies were calculated from the experimentally measured product concentrations and total charge passed during electrolysis. Each experiment was repeated independently to ensure reproducibility, and reported values represent the average of at least three independent measurements.

## References

1. Smil V. Enriching the earth: Fritz Haber, Carl Bosch, and the transformation of world food production. MIT press; 2004 Feb 27.

2. Arriola DJ, Carnahan EM, Hustad PD, Kuhlman RL, Wenzel TT. Catalytic production of olefin block copolymers via chain shuttling polymerization. Science. 2006 May 5;312(5774):714-9.

3. De Luna P, Hahn C, Higgins D, Jaffer SA, Jaramillo TF, Sargent EH. What would it take for renewably powered electrosynthesis to displace petrochemical processes?. Science. 2019 Apr 26;364(6438):eaav3506.

4. Nitopi S, Bertheussen E, Scott SB, Liu X, Engstfeld AK, Horch S, Seger B, Stephens IE, Chan K, Hahn C, Nørskov JK. Progress and perspectives of electrochemical CO2 reduction on copper in aqueous electrolyte. Chemical reviews. 2019 May 22;119(12):7610-72.

5. Garza AJ, Bell AT, Head-Gordon M. Mechanism of CO2 reduction at copper surfaces: pathways to C2 products. Acs Catalysis. 2018 Feb 2;8(2):1490-9.

6. Resasco J, Chen LD, Clark E, Tsai C, Hahn C, Jaramillo TF, Chan K, Bell AT. Promoter effects of alkali metal cations on the electrochemical reduction of carbon dioxide. Journal of the American Chemical Society. 2017 Aug 16;139(32):11277-87.

7. Lian Z, Dattila F, López N. Stability and lifetime of diffusion-trapped oxygen in oxide-derived copper CO2 reduction electrocatalysts. Nature Catalysis. 2024 Apr;7(4):401-11.

8. Kempen LH, Cheula R, Andersen M. How accurate are foundational machine learning interatomic potentials for heterogeneous catalysis?. The Journal of Chemical Physics. 2026 May 21;164(19).

9. Deng B, Choi Y, Zhong P, Riebesell J, Anand S, Li Z, Jun K, Persson KA, Ceder G. Systematic softening in universal machine learning interatomic potentials. npj Computational Materials. 2025 Jan 10;11(1):9.

10. Ringe S. The importance of a charge transfer descriptor for screening potential CO2 reduction electrocatalysts. Nature communications. 2023 May 5;14(1):2598.

11. Hori Y, Takahashi R, Yoshinami Y, Murata A. Electrochemical reduction of CO at a copper electrode. The Journal of Physical Chemistry B. 1997 Sep 4;101(36):7075-81.

12. Schouten KJ, Qin Z, Pérez Gallent E, Koper MT. Two pathways for the formation of ethylene in CO reduction on single-crystal copper electrodes. Journal of the American Chemical Society. 2012 Jun 20;134(24):9864-7.

13. Heenen HH, Shin H, Kastlunger G, Overa S, Gauthier JA, Jiao F, Chan K. The mechanism for acetate formation in electrochemical CO (2) reduction on Cu: selectivity with potential, pH, and nanostructuring. Energy & Environmental Science. 2022;15(9):3978-90.

14. Stocker S, Csányi G, Reuter K, Margraf JT. Machine learning in chemical reaction space. Nature communications. 2020 Oct 30;11(1):5505.

15. Gottweis J, Weng WH, Daryin A, Tu T, Sirkovic P, Myaskovsky A, Glowaty G, Weissenberger F, Orlandi A, Popovici D, Palepu A. Accelerating scientific discovery with Co-Scientist. Nature. 2026 May 19:1-3.

16. Ghareeb AE, Chang B, Mitchener L, Yiu A, Szostkiewicz CJ, Shved D, Gyimesi GJ, Laurent JM, Wright SM, Razzak MT, White AD. A multi-agent system for automating scientific discovery. Nature. 2026 May 19:1-3.

17. Laban P, Hayashi H, Zhou Y, Neville J. Llms get lost in multi-turn conversation. arXiv preprint arXiv:2505.06120. 2025 May 9.

18. Sclar M, Choi Y, Tsvetkov Y, Suhr A. Quantifying Language Models' Sensitivity to Spurious Features in Prompt Design or: How I learned to start worrying about prompt formatting. InInternational Conference on Learning Representations 2024 May 31 (Vol. 2024, pp. 25055-25083).

19. Nalbandyan G, Shahbazyan R, Bakhturina E. Score: Systematic consistency and robustness evaluation for large language models. InProceedings of the 2025 Conference of the Nations of the Americas Chapter of the Association for Computational Linguistics: Human Language Technologies (Volume 3: Industry Track) 2025 Apr (pp. 470-484).

20. Zhong M, Tran K, Min Y, Wang C, Wang Z, Dinh CT, De Luna P, Yu Z, Rasouli AS, Brodersen P, Sun S. Accelerated discovery of CO2 electrocatalysts using active machine learning. Nature. 2020 May 14;581(7807):178-83.

21. Sprueill HW, Edwards C, Agarwal K, Olarte MV, Sanyal U, Johnston C, Liu H, Ji H, Choudhury S. CHEMREASONER: heuristic search over a large language model's knowledge space using quantum-chemical feedback. In: Proceedings of the 41st International Conference on Machine Learning; 2024. Art. no. 1885. p. 1–24.

22. Szymanski NJ, Rendy B, Fei Y, Kumar RE, He T, Milsted D, McDermott MJ, Gallant M, Cubuk ED, Merchant A, Kim H. An autonomous laboratory for the accelerated synthesis of inorganic materials. Nature. 2023 Nov 29;624(7990):86.

23. Yang L, Lv X, Peng C, Kong S, Huang F, Tang Y, Zhang L, Zheng G. Promoting CO2 electroreduction to acetate by an amine-terminal, dendrimer-functionalized Cu catalyst. ACS Central Science. 2023 Sep 26;9(10):1905-12.

24. Wood B, Dzamba M, Fu X, Gao M, Shuaibi M, Barroso-Luque L, Abdelmaqsoud K, Gharakhanyan V, Kitchin J, Levine D, Michel K. UMA: A family of universal models for atoms. Advances in Neural Information Processing Systems. 2026 Apr 23;38:129391-427.

25. Batatia I, Kovacs DP, Simm G, Ortner C, Csányi G. MACE: Higher order equivariant message passing neural networks for fast and accurate force fields. Advances in neural information processing systems. 2022 Dec 6;35:11423-36.

26. Bilbrey, J.A., Firoz, J.S., Lee, M.S. and Choudhury, S., 2025. Uncertainty quantification for neural network potential foundation models. *npj Computational Materials*, *11*(1), p.109.

27. Bilbrey JA, Firoz JS, Allec SI, Sprueill HW, von Rueden AD, Jackson BA, Raugei S, Lee MS, Choudhury S. Assessing universal MLIP robustness with per-atom uncertainty for simulations of solid-liquid interfaces. npj Computational Materials. 2026 May 5.

28. Mazitov A, Bigi F, Kellner M, Pegolo P, Tisi D, Fraux G, Pozdnyakov S, Loche P, Ceriotti M. PET-MAD as a lightweight universal interatomic potential for advanced materials modeling. Nature Communications. 2025 Nov 27;16(1):10653.

29. Li B, Xiao J, Gao Y, Zhang JZ, Zhu T. Transition state searching accelerated by neural network potential. Journal of Chemical Information and Modeling. 2025 Feb 20;65(5):2297-303.

30. Taylor CJ, Pomberger A, Felton KC, Grainger R, Barecka M, Chamberlain TW, Bourne RA, Johnson CN, Lapkin AA. A brief introduction to chemical reaction optimization. Chemical Reviews. 2023 Feb 23;123(6):3089-126.

31. Shields BJ, Stevens J, Li J, Parasram M, Damani F, Alvarado JI, Janey JM, Adams RP, Doyle AG. Bayesian reaction optimization as a tool for chemical synthesis. Nature. 2021 Feb 4;590(7844):89-96.

32. Shinn N, Cassano F, Gopinath A, Narasimhan K, Yao S. Reflexion: Language agents with verbal reinforcement learning. Advances in neural information processing systems. 2023 Dec 15;36:8634-52.

33. Madaan A, Tandon N, Gupta P, Hallinan S, Gao L, Wiegreffe S, Alon U, Dziri N, Prabhumoye S, Yang Y, Gupta S. Self-refine: Iterative refinement with self-feedback. Advances in neural information processing systems. 2023 Dec 15;36:46534-94.

34. Yao S, Zhao J, Yu D, Du N, Shafran I, Narasimhan K, Cao Y. React: Synergizing reasoning and acting in language models. arXiv preprint arXiv:2210.03629. 2022 Oct 6.

35. Abild-Pedersen F, Greeley J, Studt F, Rossmeisl J, Munter TR, Moses PG, Skulason E, Bligaard T, Nørskov JK. Scaling properties of adsorption energies for hydrogen-containing molecules on transition-metal surfaces. Physical review letters. 2007 Jul 6;99(1):016105.

36. Stegelmann C, Andreasen A, Campbell CT. Degree of rate control: how much the energies of intermediates and transition states control rates. Journal of the American Chemical Society. 2009 Jun 17;131(23):8077-82.

37. Lum Y, Yue B, Lobaccaro P, Bell AT, Ager JW. Optimizing C–C coupling on oxide-derived copper catalysts for electrochemical CO2 reduction. The Journal of Physical Chemistry C. 2017 Jul 6;121(26):14191-203.

38. Koper MT. Theory of multiple proton–electron transfer reactions and its implications for electrocatalysis. Chemical science. 2013;4(7):2710-23.

39. GPT-5.4. Large language model. OpenAI; 2026.